# ヘテロデバイス自動オフロード時の電力使用量削減評価

山登庸次†

† NTT ネットワークサービスシステム研究所，東京都武蔵野市緑町 3-9-11
E-mail: †yoji.yamato.wa@hco.ntt.co.jp

**あらまし**　近年，少コア CPU だけでなく，GPU，FPGA，メニーコア CPU 等のヘテロなデバイスを利用したシステムが増えている．しかし，これらの利用には，CUDA 等のハードウェアを意識した技術仕様の理解が必要であり，ハードルは高い．これらの背景から，私は，プログラマーが CPU 向けに開発したソースコードを，適用される環境に応じて，自動で変換し，リソース量等を設定して，高性能，省電力で運用可能とする環境適応ソフトウェアのコンセプトを提案しており，今までに，GPU，FPGA への自動オフロードでの性能向上を検証してきた．本稿では，今まで未検証だった，自動オフロード時の電力使用量を確認し，環境適応により低電力で運用可能であることを検証する．既存アプリケーションの自動オフロード時の Watt と処理時間 Sec の積を，CPU のみで処理する場合と比較する．

# Power Reduction of Automatic Heterogeneous Device Offloading

Yoji YAMATO†

† Network Service Systems Laboratories, NTT Corporation, 3-9-11, Midori-cho, Musashino-shi, Tokyo
E-mail: †yoji.yamato.wa@hco.ntt.co.jp

**Abstract**　In recent years, utilization of heterogeneous hardware other than small core CPU such as GPU, FPGA or many core CPU is increasing. However, when using heterogeneous hardware, barriers of technical skills such as CUDA are high. Based on that, I have proposed environment-adaptive software that enables automatic conversion, configuration, and high performance and low power operation of once written code, according to the hardware to be placed. I also have verified performance improvement of automatic GPU and FPGA offloading so far. In this paper, I verify low power operation with environment adaptation by confirming power utilization after automatic offloading. I compare Watt*seconds of existing applications after automatic offloading with the case of CPU only processing.

## 1. はじめに

近年，CPU の半導体集積度が 1.5 年で 2 倍になるというムーアの法則が減速するのではないかと言われている．そのような状況から，CPU だけでなく，GPU（Graphics Processing Unit）や FPGA（Field Programmable Gate Array）等のデバイスの活用が増えている．例えば，Microsoft 社は FPGA を使って Bing の検索効率を高めるといった取り組みをしており [1]，Amazon 社は，FPGA, GPU 等をクラウド技術を用いて（例えば, [2]- [9]）インスタンスとして提供している [10]．また，IoT 機器利用のシステム（例えば, [11]- [18]) 等も増えている．

しかし，少コアの CPU 以外のデバイスをシステムで適切に活用するためには，デバイス特性を意識した設定やプログラム作成が必要であり，OpenMP（Open Multi-Processing）[19]，OpenCL（Open Computing Language）[20]，CUDA（Compute Unified Device Architecture）[21] といった知識や IoT 機器向けの組み込みシステムの知識が必要になってくるため，大半のプログラマーにとっては，スキルの壁が高い．

GPU や FPGA，メニーコア CPU 等を活用するシステムは増えていくと予想されるが，それらを活用するには，技術的壁がある．そこで，そのような壁を取り払い，少コアの CPU 以外のデバイスを十分利用できるようにするため，プログラマーが処理ロジックを記述したソフトウェアを，配置先の環境（GPU，FPGA，メニーコア CPU 等）にあわせて，適応的に変換，設定し，環境に適合した動作をさせるような，プラットフォームが求められている．

1995 年に登場した Java [22] は一度記述したコードを，別メーカーの CPU でも動作可能にし，環境適応のパラダイムシフトを起こした．しかし，移行先での性能や電力使用量は，適切であるとは限らなかった．そこで，私は，一度記述したコードを，配置先の環境に存在する GPU や FPGA，メニーコア CPU 等を利用できるように，変換，リソース設定等を自動で行い，アプリケーションを高性能かつ低電力で動作させることを目的とした，環境適応ソフトウェアを提案している．あわせて，環境適応ソフトウェアの要素として，アプリケーションコードのループ文や機能ブロックを，GPU，FPGA に自動オフロードする方式を提案し性能向上を評価している [23] [24] [25]．しかし，今までは，自動オフロード時の処理時間のみ評価しており，電力使用量については評価されていなかった．

本稿は，通常の CPU 向けプログラムを，GPU 等のデバイスにオフロードし高性能化させる際に，電力使用量も考慮に入れる方式を提案し，電力使用量低減を，既存アプリケーションのオフロードを通じて確認する．合わせて，GPU，FPGA 等の混在環境の際に，オフロード時の電力使用量も踏まえて，適切なオフロード先を自動選択する手法も提案する．

本稿の構成は以下の通りである．2 節で，既存技術を概説する．3 節で，電力使用量を考慮した GPU，FPGA への自動オフロード手法，混在環境でのオフロード先選択手法を提案する．4 節で，既存アプリケーションオフロードを通じて，GPU 自動オフロード時の電力使用量を評価する．5 節でまとめを行う．

## 2. 既存技術

### 2.1 市中技術

環境適応ソフトウェアとしては，Java がある．Java は，仮想実行環境である Java Virtual Machine により，一度記述した Java コードを再度のコンパイル不要で，異なるメーカー，異なる OS の CPU マシンで動作させている（Write Once, Run Anywhere）．しかし，移行先で性能が十分かはわからず，移行先でのデバッグや性能に関する稼働が大きかった（Write Once, Debug Everywhere）．

GPU を画像処理以外にも使う GPGPU（General Purpose GPU）（例えば [26]）を行うための環境に CUDA がある．また，GPU だけでなく，FPGA，メニーコア CPU 等のヘテロなデバイスを同様に扱うための仕様として OpenCL がある．CUDA，OpenCL は，C 言語の拡張を行いプログラムを行う形だが，プログラムの難度は高い．

CUDA や OpenCL に比べて，より簡易にヘテロデバイスを利用するため，指示行ベースで，並列処理等を行う箇所を指定して，指示行に従ってコンパイラが，GPU 等に向けて実行ファイルを作成する技術がある．仕様としては，OpenACC [27] や OpenMP 等，コンパイラとして PGI コンパイラ [28] や gcc 等がある．

CUDA，OpenCL，OpenACC，OpenMP 等の技術を用いることで，GPU や FPGA，メニーコア CPU へオフロードすることは可能になってきている．しかしオフロード自体は行えるようになっても，高速化や省電力化には課題がある．例えば，

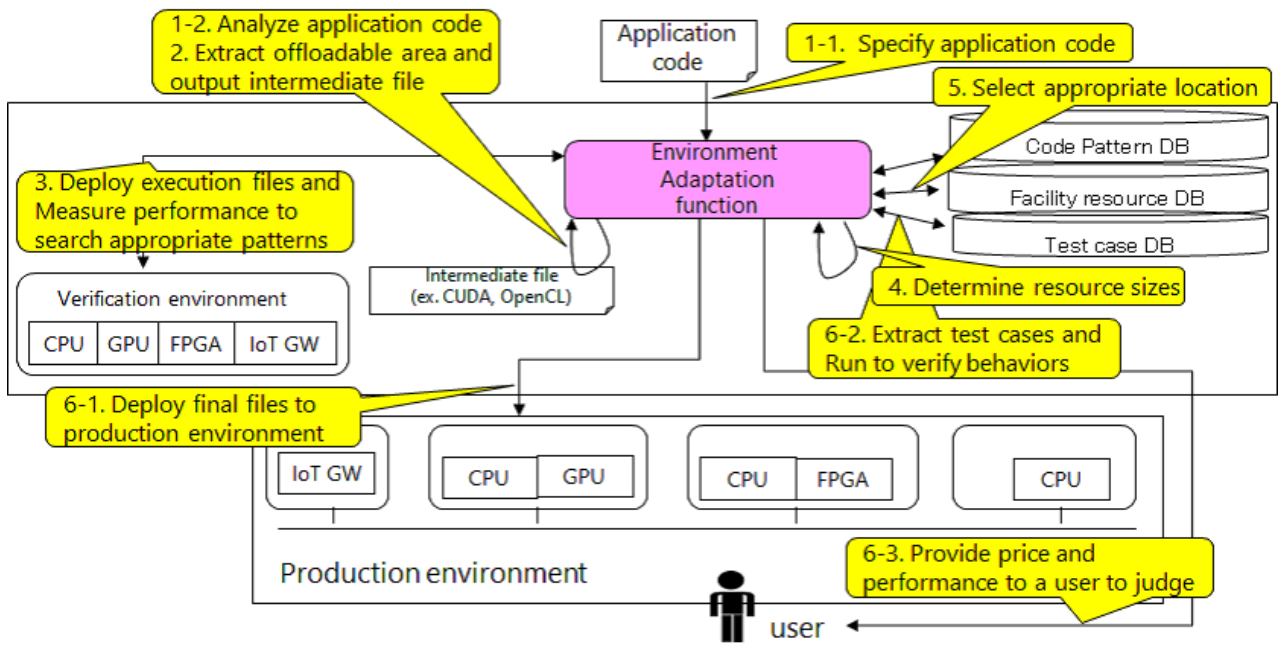

図 1 環境適応ソフトウェアのフロー

マルチコア CPU 向けに自動並列化機能を持つコンパイラとして，Intel コンパイラ [29] 等がある．これらは，ループ文中で並列処理可能な部分を抽出して，並列化している．しかし，メモリ処理等の影響で単に並列化しても性能がでないことも多い．GPU や FPGA 等で高速化する際には，OpenCL や CUDA の技術者がチューニングを繰り返している．

このため，スキルが乏しい技術者が，GPU や FPGA，メニーコア CPU を活用してアプリケーションを高速化，省電力化することは難しいし，自動並列化技術等を使う場合も並列処理箇所探索の試行錯誤等の稼働が必要だった．

### 2.2 環境適応処理のフロー

ソフトウェアの環境適応を実現するため，著者は図 1 の処理フローを提案している [31]．環境適応ソフトウェアは，環境適応機能を中心に，検証環境，商用環境，テストケース DB，コードパターン DB，設備リソース DB の機能群が連携することで動作する．

Step1 コード分析：

Step2 オフロード可能部抽出：

Step3 適切なオフロード部探索：

Step4 リソース量調整：

Step5 配置場所調整：

Step6 実行ファイル配置と動作検証：

Step7 運用中再構成：

現状を整理する．ヘテロなデバイスに対するオフロードは手動での取組みが主流である．著者は以前に環境適応ソフトウェアのコンセプトを提案し，自動オフロードを検討している．GPU 等に処理をオフロードする際に，並列処理箇所探索を自動化するため，進化計算を用いた手法を提案しているが，処理時間の短縮のみの評価であり [31]，電力使用量の削減は未評価であった．そのため，本稿では，GPU 等のオフロードデバイスに自動オフロードした際の，電力使用量削減を評価する．

## 3. 電力を考慮した GPU，FPGA への自動オフロード

著者は，環境適応ソフトウェアのコンセプトを具体化するために，これまでに，プログラムのループ文の GPU 自動オフロード，FPGA 自動オフロード，プログラムの機能ブロックの自動オフロードや，多様言語，混在環境での自動オフロードを提案

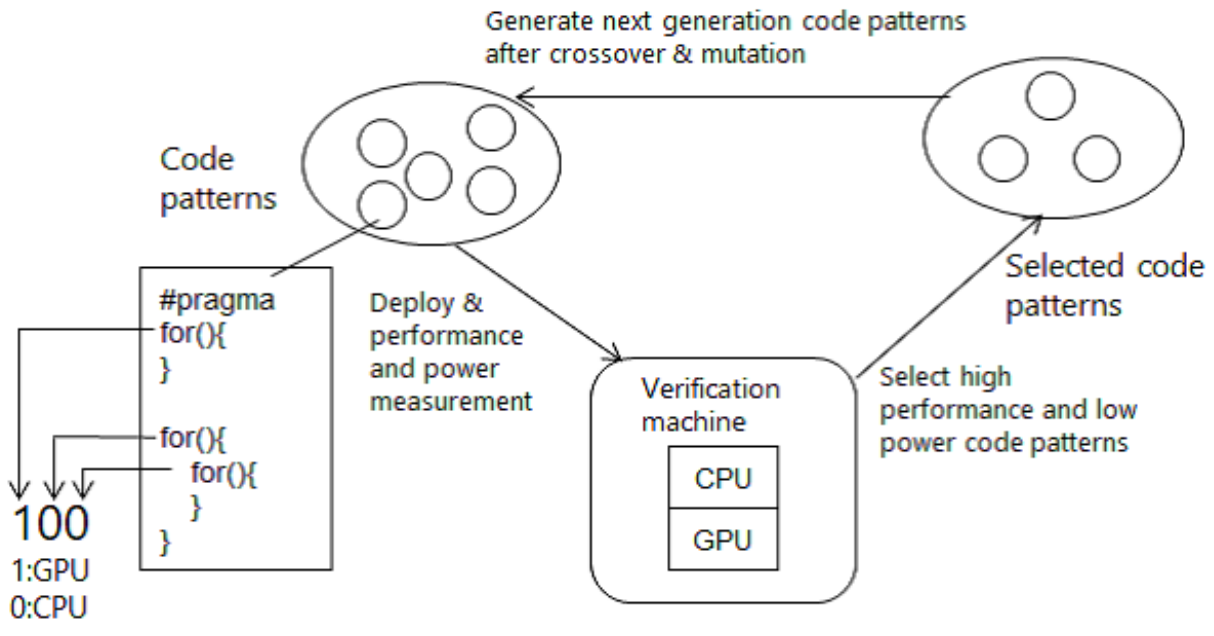

図 2　電力を考慮した GPU 自動オフロード手法

してきた．これらの要素技術検討も踏まえて，本節の，3.1，3.2 では，電力使用量を考慮したループ文の GPU，FPGA 自動オフロード技術を提案する，3.3 では，移行先が混在環境での適切なオフロード先選択技術を提案する．

### 3.1　ループ文の GPU 自動オフロード

処理時間短縮を目的にしたループ文の GPU 自動オフロード手法については，[30] [31] で提案している．

まず，共通的な課題として，コンパイラがこのループ文は GPU で並列処理できないという制限を見つけることは可能だが，このループ文は GPU の並列処理に適しているという適合性を見つけることは難しいのが現状である．実際に GPU にオフロードすることでどの程度の性能，電力消費量になるかは，実測してみないと予測は難しい．そのため，このループを GPU にオフロードするという指示を手動で行い，測定を試行錯誤することが行われている．著者はそれを踏まえ，GPU にオフロードする適切なループ文の発見を，進化計算手法である遺伝的アルゴリズム（GA）[32] で自動的に行うことを以前提案した．プログラムから，最初に並列可能ループ文のチェックを行い，並列可能ループ文群に対して，GPU 実行の際を 1，CPU 実行の際を 0 と値を置いて遺伝子化し，検証環境で反復測定し適切なパターンを探索している．ここで，検証環境測定で短時間処理できるパターンを高い適合度の遺伝子とするが，検証環境測定では電力使用量も合わせて測定し，低電力なパターンを高い適合度とする処理を新たに加える．例えば，(処理時間)$^{-1/2}$ × (電力使用量)$^{-1/2}$ のように，短処理時間，低電力使用量なほど適合度が高くなるよう設定する（図 2 参照）．

[31] では，ネストループ文中での利用変数について，ループ文を GPU 処理する際に，ネスト下位で CPU-GPU 転送が行われると下位ループ毎に転送が行われ効率的でないため，上位で CPU-GPU 転送が行われても問題ない変数については，上位で一括化転送することを提案している．ネストだけでなく，複数ファイル定義の変数も，GPU 処理と CPU 処理が入れ子にならず，CPU 処理と GPU 処理が分けられる変数については，一括化して転送する．

ループ文の GPU オフロードについては，電力使用量を適合度に含める進化計算手法と，CPU-GPU 転送の低減により，自動での高速化，低電力化を行う．

### 3.2　ループ文の FPGA 自動オフロード

著者は，ループ文の FPGA 自動オフロード手法についても，提案している [25].

FPGA で，処理時間が長時間かかる特定ループ文を FPGA にオフロードして高速化することを考えた際に，どのループをオフロードすれば高速になるかの予測は難しいため，GPU 同様検証環境で性能測定を自動で行うことを提案している．しかし，FPGA は，OpenCL をコンパイルして実機で動作させるまでに数時間以上かかるため，GPU 自動オフロードでの GA を用いて何回も反復して性能測定することは，処理時間が膨大となり行う事はできない．

そこで，FPGA にオフロードする候補のループ文を絞ってから，性能測定試行を行う形をとる．具体的には，まず，発見されたループ文に対して，ROSE フレームワーク [33] 等の算術強度分析ツールを用いて算術強度が高いループ文を抽出する．更に，gcov や gprof 等のプロファイリングツールを用いてループ回数が多いループ文も抽出する．算術強度やループ回数が多いループ文を候補として，OpenCL 化を行う．OpenCL 化時には，CPU 処理のプログラムを，カーネル（FPGA）とホスト（CPU）に，OpenCL の文法に従って分割する．算術強度やループ回数が多いオフロード候補ループ文に対して，作成した OpenCL をプレコンパイルして，リソース効率が高いループ文を見つける．これは，コンパイルの途中で，作成する Flip Flop や Look Up Table 等のリソースは分かるため，利用するリソース量が十分少ないループ文に更に絞り込む．候補ループ文が幾つか残るため，それらを用いて性能や電力使用量を実測する．選択された単ループ文に対して FPGA 実機で動作するようコンパイルして測定し，更に高速化できた単ループ文に対してはその組み合わせのパターンも作り 2 回目の測定をする．検証環境で測定された複数パターンの中で，短時間かつ低電力のパターンを解として選択する．短時間かつ低電力は，GPU の時と同様の式でスコア付けすればよい（図 3 参照）．

ループ文の FPGA 自動オフロードについては，算術強度やループ回数，リソース量を用いて候補ループ文を絞り込んでから，検証環境での測定を行い，低電力パターンの評価値を高めることで，自動での高速化，低電力化を行う．

### 3.3　混在環境での自動オフロード

著者は，GPU，FPGA，メニーコア CPU が移行先として混在している中で，高性能な移行先を選択しオフロードする技術も検討している．

検証する順番として，メニーコア CPU 向けループ文オフロード，GPU 向けループ文オフロード，FPGA 向けループ文オフロードで検証し，高性能となるパターンを探索していく．自動オフロードでは，パターンの探索は，できるだけ安価で短時間に行う事が期待される．そこで，検証時間がかかる FPGA は最後とし，前の段階で十分ユーザ要件を満足するパターンが見つかっていれば，FPGA 検証は行わない事とする．GPU とメニーコア CPU に関しては，価格的にも検証時間的にも大きな差はないが，メモリも別空間となりデバイス自体が異なる GPU に比して，メニーコア CPU の方が，通常 CPU との差は

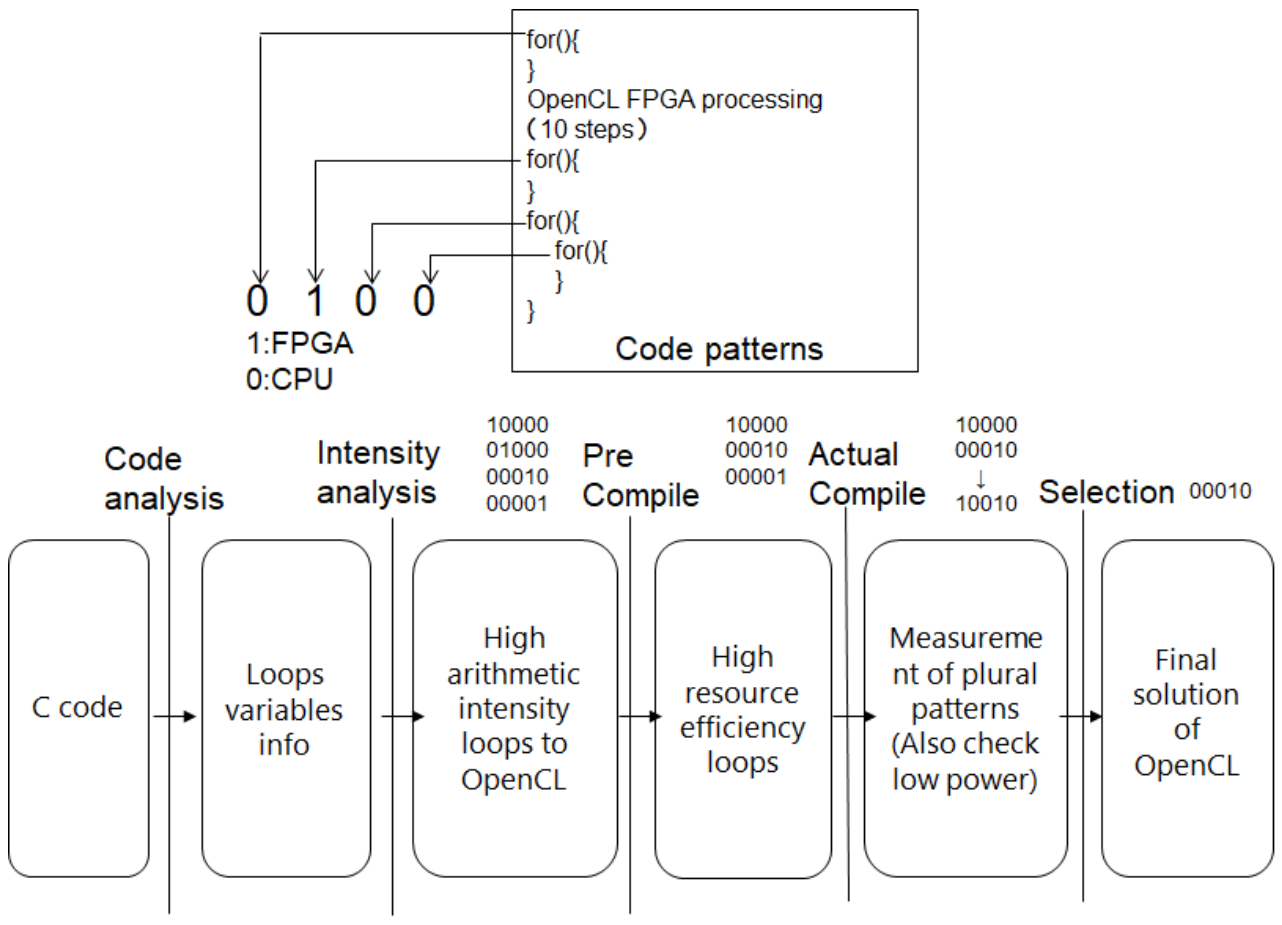

図 3　電力を考慮した FPGA 自動オフロード手法

小さいため，検証順はメニーコア CPU を先として，メニーコア CPU で十分ユーザ要件を満足するパターンが見つかっていれば，GPU 検証は行わない事とする．

ここで，3 つの移行先を検証し，高速な移行先を自動選択するのが既存方式であるが，本稿では検証環境での実測を通じて短処理時間だけでなく低電力の移行先も，自動選択の候補となるようにする．例えば，(処理時間)$^{-1/2}$ × (電力使用量)$^{-1/2}$ のように，短処理時間，低電力使用量なほどスコアが高くなるように評価式を設定すれば良い．

典型的データセンタのコストとして，ハードウェアや開発費等の初期費用が全体コストの 1/3，電力や保守等の運用コストが 1/3，サービスオーダ等のその他費用が 1/3 の例がある．この場合，例えば処理時間が 1/5 になり，CPU と GPU 合わせてもハードウェア台数が半減となれば初期費用も低減される．電力使用量半減も運用コスト低減につながる．ただし，運用コストは電力以外要素も多く，電力使用量半減が運用コスト半減になるわけでない．また，ハードウェア価格も，導入する GPU，FPGA サーバ数によりボリューム割引等があり，事業者毎に異なる．そのため，評価式は事業者毎に異なる設定とする必要がある．

このように，本稿では，処理時間だけでなく，電力使用量も考慮して，適切なオフロード先を自動選択する．

## 4. 評　　価

ループ文の GPU，FPGA への自動オフロードは [31] 等で既に評価しているため，本稿では，以前論文の実装ツールをベースに，測定パターンの評価値を定める際に，低電力な程評価値が高くなる様な改造を加えてオフロードを行い，オフロードによって，処理時間低減とともに，低電力化が出来ていることを確認する．

### 4.1 評価条件

#### 4.1.1 評価対象

評価対象は，GPU オフロードで，流体計算の姫野ベンチマークとする．

姫野ベンチマーク [34] は，非圧縮流体解析の性能測定に用いられるベンチマークソフトで，ポアソン方程式をヤコビ反復法で解いている．GPU での手動高速化に頻繁に利用されており，自動でも高速化と低電力化ができることの確認のため評価対象とする．姫野ベンチマークは C 言語や Fortran もあるが，今回は電力測定のためある程度処理時間がかかる Python を用いることとし，処理ロジックを Python で記述した．用いるデータは Large で 512*256*256 のグリッドで計算する．CPU 処理は，Python の Numpy [35] で処理され，GPU 処理は Numpy Interface で入力され GPU に処理オフロードする Cupy ライブラリ [36] を用いて処理される．

#### 4.1.2 評価手法

対象のアプリケーションのコードを入力し，移行先の GPU に対して，Clang [37] 等の解析ライブラリで認識されたループ文のオフロードを試行してオフロードパターンを決めるが，その際に，処理時間と電力使用量を測定する．最終的に定まったオフロードパターンについて，電力使用量の時間変化を取得し，オフロードせず全て通常 CPU で処理する場合に比べた電力使用量改善を確認する．

GPU での GA の条件は以下で行う．

オフロード対象ループ文数：姫野ベンチマーク 13

個体数 M：ループ文数以下とする．姫野ベンチマーク 12

世代数 T：ループ文数以下とする．姫野ベンチマーク 12

適合度：(処理時間)$^{-1/2}$ × (電力使用量)$^{-1/2}$ 処理時間と電力使用量が低い程高適合度になる．また，(-1/2) 乗とすることで，処理時間が短い特定の個体の適合度が高くなり過ぎて，探索範囲が狭くなるのを防ぐ．また，性能測定が一定時間（3 分）で終わらない際はタイムアウトさせ，処理時間 10000 秒とする．

選択：ルーレット選択．ただし，世代での最高適合度遺伝子は交叉も突然変異もせず次世代に保存するエリート保存も合わせて行う．

交叉率 Pc：0.9

突然変異率 Pm：0.05

#### 4.1.3 評価環境

利用する GPU として GeForce RTX 2080 Ti (CUDA core: 4352, Memory：GDDR6 11GB) を用いる．GPU 処理は Cupy7.8 と PyCUDA2019.1.2 を用いる．電力使用量は，GPU 搭載 PC にインストールした NVIDIA ツールの nvidia-smi コマンド [38] で GPU 電力を，s-tui コマンド [39] で CPU 電力を測定する．評価環境とスペックを図 4 に示す．

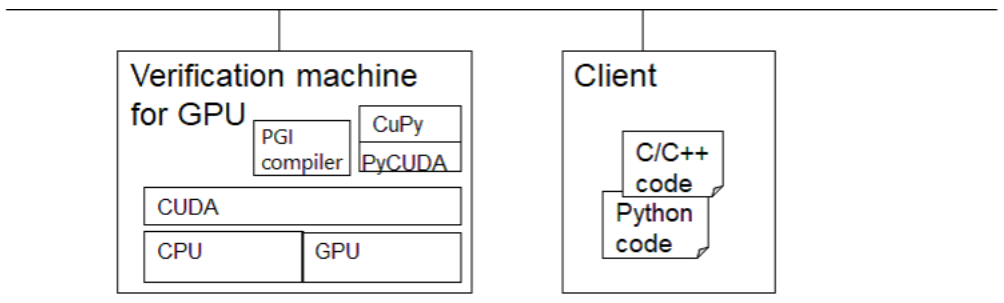

| Name | Hardware | CPU | RAM | GPU | OS | CUDA toolkit | Compiler, etc |
|---|---|---|---|---|---|---|---|
| Verification Machine for GPU | Gaming PC LEVEL-F039-LCRT2W-XYVI | AMD Ryzen Threadripper 2990WX (32 Core) | 32GB *2 | NVIDIA GeForce RTX 2080 Ti (CUDA core: 4352, Memory:GDDR6 11GB) | Ubuntu 16.04.6 LTS | 10.1 | PGI compiler 19.10, pyCUDA 2019.1.2, Cupy 7.8, Python 3 |
| Client | HP ProBook 470 G3 | Intel Core i5-6200U @2.3GHz | 8GB | | Windows 10 Pro | | |

図 4　検 証 環 境

Gaming PC (only CPU use)

| | |
|---|---|
| Watt * Sec | 4077 |
| processing time (sec) | 153 |

| time | CPU Watt |
|---|---|
| 15:04:07 | 31.2 |
| 15:04:08 | 24.2 |
| 15:04:09 | 31.0 |
| 15:04:10 | 22.9 |
| 15:04:11 | 25.7 |
| 15:04:12 | 26.9 |
| 15:04:13 | 25.0 |
| 15:04:14 | 23.8 |

execution period of Himeno benchmark

Gaming PC (CPU and GPU use)

| | |
|---|---|
| Watt * Sec | 2071 |
| processing time (sec) | 19 |

| time | CPU Watt | GPU Watt | Sum Watt |
|---|---|---|---|
| 15:03:15 | 28.9 | 27.6 | 56.5 |
| 15:03:16 | 31.5 | 27.6 | 59.1 |
| 15:03:17 | 29.7 | 27.6 | 57.3 |
| 15:03:18 | 23.4 | 41.6 | 65.0 |
| 15:03:19 | 23.9 | 66.7 | 90.6 |
| 15:03:20 | 24.2 | 81.9 | 106.1 |
| 15:03:21 | 22.3 | 91.3 | 113.6 |
| 15:03:22 | 25.1 | 91.1 | 116.2 |

図 5　GPU オフロード時電力使用量（姫野ベンチマーク）

### 4.2　結果と考察

図 5 は GPU に姫野ベンチマークをオフロードした際の，Watt と時間を記載している．図より，全て CPU 処理する場合と比較して，処理時間は 153 秒から 19 秒に短縮されているが，電力は CPU のみの場合の 27W 程度から CPU と GPU が使われ 109W 程度となっていることがわかる．その結果，Watt*sec は，CPU のみ処理の場合の 4080Watt*sec から，GPU にオフロードした場合で 1/2 の 2070Watt*sec となった事が分かる．

IoT 等で多くのユーザが利用すると想定されるアプリケーションとして，流体計算の姫野ベンチマークの高速化と低電力化を確認した．GPU にオフロードした際は，Watt 自体は増えるが，短時間化の効果が高く全体として Watt*sec が減っている．FPGA オフロードの際の電力使用量は今回測定にはないが，GPU，FPGA 混在環境ならば，オフロード時の性能と電力から適切なオフロード先を選択する．

## 5.　ま　と　め

本稿では，私が提案している，ソフトウェアを配置先環境に合わせて自動適応させ GPU，FPGA 等を適切に利用して，アプリケーションを高性能，低電力で運用するための環境適応ソフトウェアの要素として，電力使用量を考慮したオフロード手法を提案し評価した．

GPU，FPGA 自動オフロード時に検証環境で実測する際に，処理時間とともに電力使用量を取得し，短時間かつ低電力なパターンを高い適合度とすることで，自動でのコード変換に低電力化を盛り込む．GPU，FPGA 等が混在している場合は，単体の移行先移行を試行し，短時間かつ低電力な移行先を選択するようにすることで，自動選択を行う．流体計算の GPU 自動オフロードを通じて，高性能化と共に低電力化を確認し，方式の有効性を確認した．

今後は，より多くのアプリケーションでの電力使用量低減を検証する．また，事業者のコスト構造の具体例を参考に，短処理時間化と低電力化の評価式を検討する．

### 文　献